\documentclass{ws-ijmpcs}

\begin{document}

\markboth{C. Hanhart}
{Heavy Exotic Mesons --- Theory}

%
\catchline{}{}{}{}{}
%

\title{Theory Concepts for Heavy Exotic Mesons}

\author{C. Hanhart}

\address{Forschungszentrum J\"ulich, Institute for
           Advanced Simulation, Institut f\"ur Kernphysik and
           J\"ulich Center for Hadron Physics,\\ D-52425 J\"ulich, Germany\\  c.hanhart@fz-juelich.de}

\maketitle

\begin{history}
\published{Day Month Year}
\end{history}

\begin{abstract}
Some of the currently most popular conjectures for the structure of the recently discovered
heavy mesons that do not find a place in the quark model quarkonium spectrum are sketched.
Furthermore,
some observables are identified that should allow one to identify the most prominent components
of individual states.   
\end{abstract}

\section{Introduction}	

The observation of the $X(3872)$ in 2003\cite{Choi:2003ue} initiated a renaissance of heavy meson spectroscopy, 
since the properties of this particle were in conflict with the predictions of the quark model, which was
 until then
extremely successful  in the heavy quark sector. Since then
a large number of additional candidates of exotic states was discovered. What is intriguing in this context is
the fact that below the first open flavor threshold the quark model continues to provide an extremely successful
description of the properties of the quarkonium states --- even those discovered after 2003 --- while all heavy\footnote{Here
the notion of 'heavy exotics' is used for exotics that contain a heavy quark--anti-quark pair.} exotics
reside above this threshold. For long there was  the expectation by some that with some adjustment of the
quark model it is possible to describe the $X(3872)$ as a realization of the $\chi_{c1}(2P)$ quark model state, however,
with the discovery of charged states with prominent decays into heavy quarkonia\cite{Belle:2011aa} it became apparent that 
meson states beyond the most simple realization of the quark model exist. For general reviews about the
heavy exotics we refer to Refs.~\refcite{Brambilla:2014jmp,Lebed:2016hpi,Olsen:2017bmm} --- reviews with specific emphasis on certain aspects are quoted
in the sections below.


The focus of this presentation is on some representative phenomenological approaches that can be found in
the literature aiming at a better understanding of the newly discovered states. In general one can distinguish
two classes of states beyond the naive quark model, namely those where gluonic excitations contribute to
the quantum numbers, like glueballs and hybrids, and multi-quark states. In this presentation I would like to
focus on the latter class. To be concrete, I will compare the phenomenology of and predictions for tetraquarks, hadro-quarkonia and 
hadronic molecules. In simple words these three may be distinguished by the way the heavy and light quarks
arrange themselves within a given hadron:  The fundamental building blocks of tetraquarks are colored
heavy light diquarks and anti-diquarks, those of hadro-quarkonia are compact, colorless quarkonia, surrounded
by a light quark cloud and those of hadronic molecules are pairs of heavy open flavor mesons.
The phenomenological
implications of these structures are discussed in some detail in the following sections.

\section{Tetraquarks}

The first detailed study of tetraquark structures dates back to the works of Jaffe employing the MIT-bag model\cite{Jaffe:1976ig}.
Already at that time the notion of the $[\bar 3]$ diquarks as `good diquarks' was initiated based on the observation that
in a pertrubative treatment this is the attractive channel --- most of the following works therefore included only the good
diquarks.
While in Ref.~\refcite{Jaffe:1976ig} the focus was on the light quark sector, in 2003, shortly after the discovery of the $X(3872)$ 
in Ref.~\refcite{Maiani:2004vq} the picture was adapted to the quarkonium sector. While in this original work the spin-spin interaction
was dominated by that between the diquarks, in line with expectations from the heavy quark symmetry, it
was realized in Ref.~\refcite{Maiani:2014aja} that the $Z_c$ spectrum can be described only, if the spin-spin interaction acts
predominantly within the diquarks. In this work it is also shown that this unusual assumption also helps to describe
some decay phenomenology --- in particular the transition $Y(4260)\to X(3872)\gamma$.
For a recent review of multiquark states with emphasis on tetraquarks we refer to Ref.~\refcite{Esposito:2016noz}.

The approach starts from identifying the most relevant diquark--anti-diquark interactions contributing to
the mass of a tetraquark as\cite{Maiani:2014aja}
 \begin{eqnarray}\nonumber
  M&=&\hat M_{00}+\frac{B_c}{2}\vec L\, ^2-2a \vec L\cdot \vec S+ \, 2\kappa_{cq} \left[\vec s_c\cdot \vec s_q + \vec s_{\bar c}\cdot \vec s_{\bar q} \right] \\ \nonumber
&=& M_{00}+B_c\frac{L(L+1)}{2}+a[L(L+1)+S(S+1)-J(J+1)]\\ 
& & \hspace{2.cm}+ \, \kappa_{cq} \left[s(s+1)+\bar{s}(\bar{s}+1)-3\right] \ ,
\label{Htetra}
 \end{eqnarray}
 where in the last step it was used that the Hamilonian acts on diquarks and anti-diquarks with definite spin $s$ and $\bar s$, respectively, coupled to a given
 angular momentum $L$ and total spin $S$ combined to some total angular momentum $J$. 

To fix the parameters one may start with the $L=0$ sector which for tetraquarks refers to even parity states. Then 
one is faced with three established states, namely $X(3872)$ --- an isoscalar with $J^{PC}=1^{++}$, and
$Z_c(3900)$ and $Z_c(4020)$, both isovectors with $J^{PC}=1^{+-}$ which allow one to fix $\kappa_{cp}$ as
well as $M_{00}$. Before one can proceed to fix the parameters that enter for $L=1$, which need to be fixed
from negative parity states, one  needs to decide which states to include in the fit --- at present the spectrum
of $1^{--}$ states is not established. For example, in Ref.~\refcite{Ali:2017wsf} the fit--scheme of
Ref.~\refcite{Maiani:2014aja}, which included $Y(4008)$, $Y(4260)$, $Y(4360)$ and $Y(4660)$,
was contrasted with an alternative scheme including $Y(4220)$, $Y(4330)$\footnote{This state is often
referred to as $Y(4320)$.}, $Y(4390)$ and $Y(4660)$
--- in line with the most recent measurements of BESIII\cite{Ablikim:2016qzw,talkChengping}.
Furthermore the interaction was extended to allow for a spin-spin tensor force in addition to the terms
mentioned above. 
In any case the analysis calls for 4 exotic states in the $1^{--}$ sector below $4700$ MeV for the
parameters to be in line with other systems. It remains to be seen how many of the states claimed at present
get established eventually. Implications of Eq.~(\ref{Htetra}) for other quantum numbers are discussed
in Ref.~\refcite{Cleven:2015era}.

It should also not remain unmentioned that there are various problems with the tetraquark picture as
presented. First of all it predicts a lot more states than observed: There are not only those that follow
directly from Eq.~(\ref{Htetra}). The spectrum is on top doubled by the fact that all states should appear near
degenerate in both the isoscalar and the isovector channel in complete analogy to the proximity in masses of
$\rho$ and $\omega$. In addition it is also not clear yet, if just looking at good diquarks is appropriate\cite{Richard:2017vry}.

Recently there appeared growing interest into $QQ\bar q\bar q$ tetraquarks. There exist now studies
from QCD sum rules\cite{Du:2012wp}, lattice QCD\cite{Francis:2016hui} as well as phenomenology\cite{Karliner:2017qjm,Eichten:2017ffp}.
Especially the last works employ the observation of doubly heavy baryons to make predictions for doubly heavy
tetraquarks. The connection between these systems might be most compactly collected into the formula\cite{Eichten:2017ffp}
\begin{equation}
{m(QQ\bar q\bar q) - m(QQ q) \simeq m(\bar Q\bar q\bar q)-m(\bar Q q)} \ ,
\label{doublyheavy}
\end{equation}
which is based on a proposed quark-diquark symmetry\cite{Savage:1990di}. This symmetry is realized in nature,
if  heavy diquarks form compact substructures in hadrons, for this would allow one to perform a systematic expansion
in $r_{QQ}/r_{qq}$, where $r_{qq}$ denotes the size of the light quark cloud that may be estimated as $1/\Lambda_{\rm QCD}$.
As soon as $r_{QQ}/r_{qq}$ is a small parameter, one may safely assume the $QQ$ diquark to be in the color--anti-triplet 
configuration, since for heavy quarks the $QQ$ interaction should be largely governed by the one gluon exchange which
is attractive only in this channel. Then Eq.~(\ref{doublyheavy}) follows naturally. Especially since recently the first
doubly heavy baryon was found experimentally\cite{Aaij:2017ueg},
the conclusion might well be inverted:
If the pattern of Eq.~(\ref{doublyheavy}) were not realized in nature, it would tell us that QCD does not favor 
doubly heavy compact diquarks. Therefore the experimental search for the mentioned tetraquark structures should be 
performed with high priority at, e.g. Belle and LHCb.
The above mentioned studies find typically a deeply bound $bb\bar u\bar d$ system with $J^P=1^+$ 100-200
MeV below the $\eta_b\eta_b$ threshold.

\section{Hadroquarkonia}

The physical picture underlying  hadroquarkonia is that of a compact quarkonium core surrounded
by a light quark cloud sticking together thanks to the QCD analogue of the van der Waals force\cite{Dubynskiy:2008mq}.  
In this approach the decay of, say, the $Y(4260)$ viewed as a $J/\psi$ core
with an isoscalar pion cloud happens simply by separation of core and cloud.
This provides a natural explanation, why states like $Y(4260)$ and $Y(4360)$ are seen in
$e^+e^-\to J/\psi\pi\pi$ and $e^+e^-\to \psi(2S)\pi\pi$, respectively, but not in $D^{(*)}D^{(*)}$ as
expected for $\bar cc$ states. 
In order to explain the observation of both $Y(4260)$ and $Y(4360)$ in the $h_c(1P)\pi\pi$ final
state, where the transition to the spin 0 charmonium in the final state seems to suggest a significant
amount of heavy-quark spin symmetry violation, it was proposed in Ref.~\refcite{Li:2013ssa} that $Y(4260)$ and $Y(4360)$
emerge from a mixing of a charmonium with a spin 1 core (dominated by $\psi(2S)$) and one
with a spin 0 core (predominantly $h_c(1P)$).

Predictions derived from the hadro--quarkonium approach follow most straightforwardly for the
 partner states derived employing heavy quark spin symmetry. For example, when the $Z_c(3900)$
 is assumed to be a hadro--charmonium composed of a $J/\psi$ core supplemented with a
 light quark cloud carrying the quantum numbers of the pion, then there must be a spin partner
 state composed of an $\eta_c(1S)$ core with the same cloud\cite{Voloshin:2013dpa}. 
 In particular, this partner states baptized $W_c$ 
should be lighter than the $Z_c(3900)$ according to
\begin{equation}
M_{W_c} = M_{Z_c(3900)}-M_{J/\psi}+M_{\eta_c} \ .
\end{equation}
On the contrary, if the $Z_c(3900)$ is a molecular state composed of $D\bar D^*$, then the
spin partner should be heavier than the $Z_c$ by the $D^*$-$D$ mass difference\cite{Voloshin:2013dpa}. Therefore, with the
discovery of an isovector state near the $D^*\bar D^*$ threshold at BESIII\cite{Ablikim:2013wzq}
the hadro-quarkonium picture is ruled out for the $Z_c(3900)$. 

When a similar logic to identify spin partners  is applied to the scenario outlined above 
for $Y(4260)$ and $Y(4360)$, in total 4 spin partners emerge with the distinctive feature
that there emerges a relatively light  $\eta_c(4140)$ absent from the phenomenology for
 the other exotic scenarios\cite{Cleven:2015era}. A more detailed comparison of
 features deduced from the hadro-charmonium picture and those deduced from the
 molecular picture for $Y(4260)$ can be found in Ref.~\refcite{Wang:2013kra}.

\section{Hadronic Molecules}

Hadronic molecules are bound states of two (or more) color neutral states in close analogy
to nuclei. The binding force typically comprises both short ranged interactions as well
as the longer ranged exchange of the relatively light Goldstone bosons --- especially the
pion~\footnote{Note that the pion exchange potential with derivative couplings 
is field theoretically consistent only when accompanied by 
 short ranged interactions\cite{Baru:2015nea}.}, which is regarded by many as 
crucial whenever it is allowed to contribute at tree level.
Based on meson exchange models deuteron--like states composed of two heavy mesons were predicted 
already in Refs.~\refcite{Voloshin:1976ap,Tornqvist:1993ng}.
For a recent review on molecular states we refer to Ref.~\refcite{Guo:2017jvc}.

It is well known from nuclear physics that binding is in general the strongest in $S$--waves. Accordingly
we here focus on $S$--wave systems. Furthermore, molecular states can only be formed
from narrow hadrons --- broad constituents would necessarily lead to broad bound states\cite{Filin:2010se}; the same statement
from a different angle may be phrased as: The constituents must live long enough to allow the bound
state to form\cite{Guo:2011dd}.

For shallow bound states Weinberg developed a criterion that allows one to extract the probability
to find the molecular component in the wave function of this state, $(1-\lambda^2)$, directly from the residue at the
bound state pole (which is the square of the effective coupling of the given state to the relevant 
continuum channel)\cite{Weinberg:1965zz}:
\begin{equation}
\frac{g^2_{\rm eff.}}{4\pi} = \frac{4M^2\gamma}{\mu}(1-\lambda^2) \ ,
\label{weincrit}
\end{equation}
where $M$ denotes the mass of the bound state, $\mu=m_1m_2/(m_1+m_2)$ the reduced mass of the two constituents
with mass $m_1$ and $m_2$, respectively,
and $\gamma=\sqrt{2\mu (m_1+m_2-M)}$ is the binding momentum. In Ref.~\refcite{Baru:2003qq} it was
shown that this relation can still be used even in the presence of remote inelastic channels. 
Eq.~(\ref{weincrit}) acquires corrections of the order of $\gamma R$, where $R$ denotes the range of
forces. Thus, those corrections can get quite large especially for systems with heavy mesons. However, 
the spirit of Eq.~(\ref{weincrit}) is easily generalized: $(i)$ Transitions with the constituents in the final state
are enhanced; $(ii)$ loops that contain the constituents appear at leading order.
 Then, for all observables where those loops are convergent, the 
 molecular nature of the states is expected to leave an imprint in these
observables~\cite{Cleven:2013sq}. However, if these loops are divergent, the observable is sensitive to the short ranged structure
of the wave function that is not under control in the effective field theory. This calls for a counter
term at leading order --- examples of observables of this class are radiative decays of $X(3872)$\cite{Guo:2014taa} 
or molecule production at high $p_T$\cite{Albaladejo:2017blx}. 

\begin{figure}[t!]
\centering
\includegraphics[width=13cm,clip]{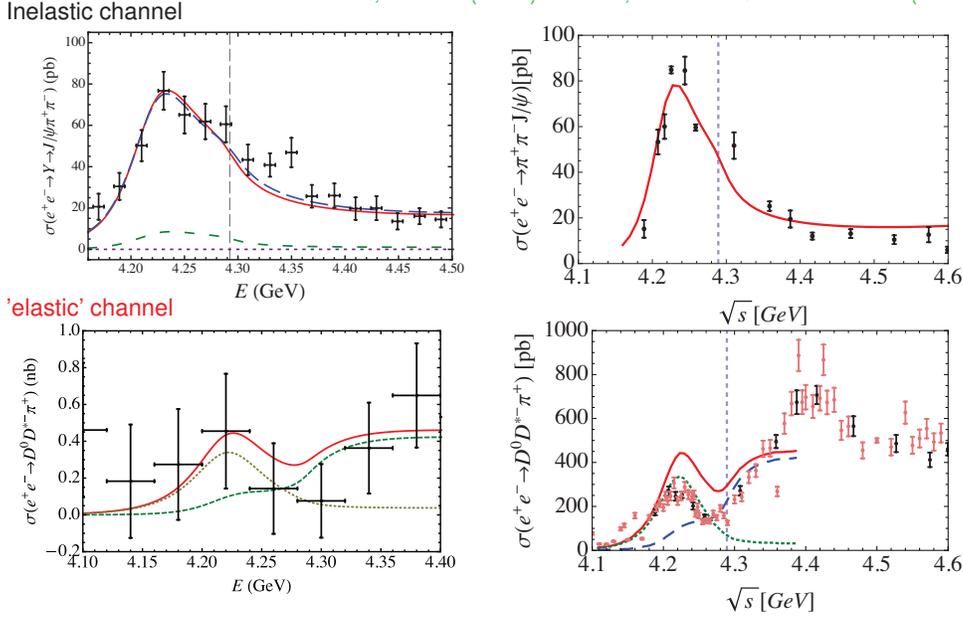}
\caption{The lineshapes of $Y(4260)$ as seen in both the $J/\psi\pi \pi$ (first line) and the $D^*\bar D\pi$ (second line)
final states. The left panels show the data as available 2014\protect\cite{Liu:2013dau,Pakhlova:2009jv}, when the parameters of the molecular model described
in the text were fixed. In the upper panel left the dotted, green short-dashed, blue long-dashed and solid line refer
to the background, the $Z_c$, the box and the full contribution. In the lower left panel the dotted, dashed and solid
line refer to the $D^*\pi$ $S$-wave, the $D^*\pi$ D-wave and the full contribution. For details we refer to Ref.~\protect\cite{Cleven:2013mka}.
The right panels show the previously fixed line shapes compared to the most recent data
of Ref.~\protect\refcite{Ablikim:2016qzw}
and \protect\refcite{talkChengping} for the $J/\psi \pi\pi$ and $D^*\bar D\pi$ channels, respectively. Note that the latter
data set is still preliminary.
\label{Yspectra} }      
\end{figure}

Natural observables sensitive to the molecular nature are the line shapes of a given state. Those are especially
non-trivial when the constituents are unstable as was first observed in Ref.~\refcite{Braaten:2007dw} (see also Ref.~\refcite{Hanhart:2010wh}).
This is illustrated in Fig.~\ref{Yspectra}: Shown are the experimental line shapes of $Y(4260)$ in the $J/\psi\pi\pi$ channel (first row)
and the $D^*\pi \bar D$ channel (second row) compared to those that emerge from a calculation
assuming a $D_1(2420)\bar D$ molecular structure for this $1^{--}$ state\cite{Cleven:2013mka} as
proposed in Ref.~\refcite{Wang:2013cya}. Since $D^*\pi$ in a D-wave is the main decay channel
of the $D_1(2420)$ the latter channel reflects the decay of the molecule into its constituents, if indeed the
state is of molecular nature. The model\cite{Cleven:2013mka} was fit to the data available
in 2013 --- especially the spectrum of Ref.~\refcite{Liu:2013dau}, shown in the upper left panel of Fig.~\ref{Yspectra}, as
well as some differential spectra provided by BESIII~\cite{Ablikim:2013xfr}. In particular the angular
distributions published there called for some small admixture of the $D^*\pi$ $S$--wave, as shown by the dotted line 
in the figure --- the dashed line shows the $D$-wave and the solid line the sum of both.
Note that the line shape of Ref.~\refcite{Cleven:2013mka} in $J/\psi\pi\pi$ peaks near 4220 MeV and not at
4260 MeV. On the other hand a mass of 4260 MeV was extracted from these data using a symmetric Breit-Wigner distribution.
 The data for $e^+e^-\to D^*\pi\bar D$ existing
at the time\cite{Pakhlova:2009jv}, shown in the lower left panel, was not included in the fit --- the distribution emerged as a prediction.
It is therefore important to note that in the molecular picture very non--trivial line shapes emerge naturally for
$Y(4260)$. 

In the right column the most recent data in the $Y(4260)$ mass range for the $J/\psi\pi\pi$~\cite{Ablikim:2016qzw}
and the $D^*\pi\bar D$\cite{talkChengping} channel are shown together with the same  curves already fixed
in 2013. Obviously the gross features of the data are nicely reproduced in both channels with
just one single particle state included\cite{Cleven:2013mka,Qin:2016spb}. On the contrary, based on analyses that again
involved only Breit-Wigner distributions, the data shown were claimed to contain
in addition to a pole at 4222 MeV and a higher lying one\cite{Ablikim:2016qzw,Gao:2017sqa}.

Besides the line shapes also the quantitative implications of spin symmetry violations turn out to be quite specific
for the underlying dynamics. For the case of the spin partners of heavy meson molecules this is worked
out in some detail in Refs.~\refcite{Bondar:2011ev,Voloshin:2011qa,Mehen:2011yh,Nieves:2012tt,Baru:2016iwj,Baru:2017gwo}.

\section{Summary and Conclusions}

We are facing exciting times in heavy meson spectroscopy. As outlined in the presentation there is
reason to hope that the upcoming high precision data from BESIII, LHCb, Belle and BaBar as well
as from the future detectors BelleII and PANDA will allow us to understand the spectrum of the
newly discovered heavy meson exotics. What is especially needed to discriminate between different
theoretical pictures is information about resonances in other channels as well as detailed measurements
of line shapes. 

Before closing a disclaimer is in order: Clearly in reality the different configurations will mix. On top
of what is described here there can furthermore occur a mixing with regular mesons. Theoretical
investigations of these mixings are still in their infancy\cite{Rupp:2015taa,Cincioglu:2016fkm,Hammer:2016prh}. However, I regard
it as realistic --- a judgement that might well be driven by desire --- that the gross features of the 
properties of the exotics can be mapped onto one of the scenarios outlined in this presentation.
Those studies will then eventually reveal, how (or if at all) QCD clusters its constituents
into multi-quark states providing crucial insights in the inner workings of the strong interaction.

\section*{Acknowledgments}

The author is grateful for the various enlightening collaborations with Vadim Baru, Martin Cleven, Feng-Kun Guo, Yulia Kalashnikova, 
 Ulf-G. Mei\ss ner, Alexey
Nefediev, Qian Wang, 
Qiang Zhao and Bing-Song Zou that lead to the results presented here.
This work is supported in part by the DFG and the NSFC through funds provided to the Sino-German CRC 110 ``Symmetries
and the Emergence of Structure
in QCD''.


\end{document}